\documentclass[twocolumn]{aastex62}

\graphicspath{{./}{figures/}}

\usepackage{color}
\usepackage{ulem}

\received{2019 January 8}
\revised{2019 March 22}
\accepted{2019 March 30}

\shorttitle{[\ion{O}{4}] 26 MICRON AND 12 MICRON DIAGNOSTICS FOR "BURIED" AGN}
\shortauthors{Yamada et al.}

\begin{document}

\title{LUMINOSITY RATIO BETWEEN [\ion{O}{4}] 25.89 $\mu$m LINE AND NUCLEAR 12 $\mu$m CONTINUUM \\
AS DIAGNOSTICS FOR ``BURIED'' AGN}

\author{Satoshi Yamada}
\affiliation{Department of Astronomy, Kyoto University, Kitashirakawa-Oiwake-cho, Sakyo-ku, Kyoto 606-8502, Japan; styamada@kusastro.kyoto-u.ac.jp}

\author{Yoshihiro Ueda}
\affiliation{Department of Astronomy, Kyoto University, Kitashirakawa-Oiwake-cho, Sakyo-ku, Kyoto 606-8502, Japan; styamada@kusastro.kyoto-u.ac.jp}

\author{Atsushi Tanimoto}
\affiliation{Department of Astronomy, Kyoto University, Kitashirakawa-Oiwake-cho, Sakyo-ku, Kyoto 606-8502, Japan; styamada@kusastro.kyoto-u.ac.jp}

\author{Taiki Kawamuro}
\affiliation{National Astronomical Observatory of Japan, Osawa, Mitaka, Tokyo 181-8588, Japan}

\author{Masatoshi Imanishi}
\affiliation{National Astronomical Observatory of Japan, Osawa, Mitaka, Tokyo 181-8588, Japan}
\affiliation{Department of Astronomical Science, Graduate University for Advanced Studies (SOKENDAI), 2-21-1 Osawa, Mitaka, Tokyo 181-8588, Japan}

\author{Yoshiki Toba}
\affiliation{Department of Astronomy, Kyoto University, Kitashirakawa-Oiwake-cho, Sakyo-ku, Kyoto 606-8502, Japan; styamada@kusastro.kyoto-u.ac.jp}
\affiliation{Academia Sinica Institute of Astronomy and Astrophysics, 11F of Astronomy-Mathematics Building, AS/NTU, No.1, Section 4, Roosevelt Road, Taipei 10617, Taiwan}

\begin{abstract}
We propose new diagnostics that utilize the [\ion{O}{4}] 25.89 $\mu$m and
nuclear (subarcsecond scale) 12~$\mu$m luminosity ratio for identifying
whether an AGN is deeply ``buried''
in their surrounding material. Utilizing a sample
of 16 absorbed AGNs at redshifts lower than 0.03 in the \textit{Swift}/BAT catalog observed with
\textit{Suzaku}, we find that AGNs with small scattering fractions ($<$0.5\%)
tend to show weaker [\ion{O}{4}]-to-12 $\mu$m luminosity ratios
than the average of Seyfert 2 galaxies. This suggests that this ratio is
a good indicator for identifying buried AGNs. Then, we apply this
criterion to 23 local ultra/luminous infrared galaxies (U/LIRGs) in
various merger stages hosting AGNs. We find that AGNs in most of 
mid- to late-stage mergers are buried, while those
in earlier stage ones (including non-merger) are not.
This result suggests that the fraction of buried
AGNs in U/LIRGs increases as the galaxy-galaxy interaction becomes more
significant.

\end{abstract}

\keywords{galaxies: active --- galaxies: nuclei --- X-rays: galaxies --- infrared: galaxies}

\section{Introduction} \label{sec:intro}

Luminous infrared galaxies (LIRGs; $10^{11} L_{\odot} \leqq L_{\rm IR}$(8--1000\
$\mu$m) $< 10^{12} L_{\odot} $) and ultraluminous infrared galaxies
(ULIRGs; $L_{\rm IR} \geqq 10^{12} L_{\odot}$) have large bolometric
luminosities that are radiated mostly as infrared dust emission 
\citep{Sanders1988,Sanders1996}. This indicates that their
powerful energy sources, starburst, and/or active galactic nuclei (AGNs)
are hidden behind gas and dust.  The contribution of U/LIRGs to
the total infrared luminosity density increases rapidly with redshift
\citep[e.g.,][]{Goto2010}.
Therefore, the study of U/LIRGs is important to understand starburst-AGN
connection, and the history of dust-obscured starburst and supermassive
black hole (SMBH) growth in the universe.

The majority of U/LIRGs in the local universe are known to be merging
systems of gas-rich disk galaxies \citep[e.g.,][]{Veilleux2002,Kartaltepe2010}.
The merger is considered to be a key mechanism
to funnel material from the kpc- to the pc-scale environment of SMBHs.
It is predicted that during the final phase of a merger such as in
U/LIRGs, rapid accretion onto the SMBHs takes place when the nucleus is
deeply enshrouded by gas and dust \citep[e.g.,][]{Hopkins2006}.  Such
late-stage mergers are likely become ``buried AGNs'',  
where even the direction of the lowest dust column-density
can be opaque to the ionizing UV photons (i.e., the
covering fraction of obscuring material, which we conventionally call
``the torus'' in this paper, is close to unity) and thereby the narrow
line region (NLR) is less developed compared with normal AGNs \citep[e.g.,][]{Imanishi2008}.  Thus, in
order to test the scenario of merger-driven SMBH growth, it is important
to identify buried AGNs and compare their fractions among different
merger stages. 
In fact, hard X-ray observations of local U/LIRGs have
revealed that AGNs in later stage mergers have larger amount of material
around the SMBH \citep{Ricci2017b}. However, since X-ray data are not
always available and have limitations (see below), it is quite useful to
establish other methodology for identifying buried AGNs.

Broadband X-ray spectra
covering from sub keV to several tens of keV
provide useful diagnostics to identify  
whether an obscured AGN is deeply buried, 
as long as it is not heavily Compton thick (i.e., log$N_{\rm
H}/$cm$^{-2} >$ 25). 
A typical X-ray spectrum of an obscured AGN consists of 
an absorbed direct power-law component, 
its reflected components from dense material such as the accretion disk and torus accompanied by fluorescent lines, 
and a scattered component by NLR gas, which is observable as a weak unabsorbed
component in the soft X-ray band.
It is expected that the intensity of X-rays scattered by the NLR gas
relative to that of the direct component (scattering fraction; $f_{\rm scat}$) decreases
with the torus covering fraction. Indeed, X-ray observations have
discovered ``low X-ray-scattering AGNs'', which are good candidates of
AGNs deeply buried in geometrically thick tori \citep[e.g.,][]{Ueda2007,Ueda2015}. 
To apply this method, however, we need X-ray spectra with
sufficiently high quality. A potential problem for U/LIRGs is possible
contamination from hot gas 
and X-ray binaries
in the star forming regions, which makes it difficult to correctly estimate the flux
of scattered AGN X-rays.

Another approach is to use the luminosities of spectral lines from
AGN-excited ions in the NLR, which are proportional to the NLR size
(hence the opening angle of the torus) and AGN luminosity at
wavelengths responsible for the ionization. In particular, mid-infrared (MIR)
lines from ions with highly ionization potentials, such as [\ion{O}{4}] 25.89
$\mu$m (54.9 eV), are quite useful because they are less subject to extinction by
dust in the host galaxy and are less contaminated by star formation
activities than widely-used optical lines such as [\ion{O}{3}]
$\lambda$5007. For instance, the ratio between the [\ion{O}{4}] 25.89~$\mu$m
and X-ray (2--10 keV) luminosities could be a good measure for
identifying whether an AGN is buried. 
In fact, \citet{Kawamuro2016a} found a good
correlation between this ratio and the X-ray scattering fraction, using
a sample of hard X-ray selected AGNs. A disadvantage of using the 
$L_{\rm [O\ IV]}/L_{\rm X}$ ratio is its coupling with the UV to X-ray spectral energy
distribution (SED), because the [\ion{O}{4}] 25.89 $\mu$m luminosity is
proportional to that in the UV (not X-ray) band. It has been suggested
that the X-ray to bolometric luminosity ratio of an AGN depends on the
Eddington ratio \citep[e.g.,][]{Vasudevan2007,Toba2019}. This leads to
degeneracy, in particular for AGNs in U/LIRGs, whose Eddington ratios
may be much larger than in normal Seyfert galaxies \citep[see e.g.,][]{Oda2017}.

In this paper, we propose the luminosity ratio between the [\ion{O}{4}] 25.89
$\mu$m line and ``nuclear'' (subarcsecond scale) 12 $\mu$m 
continuum as new diagnostics for
identifying whether an AGN is deeply buried.
Assuming that the MIR luminosity originates
from hot dust heated by the AGN \citep[e.g.,][]{Gandhi2009}, it should basically be
proportional to the bolometric AGN luminosity (dominated by the UV
luminosity) times the torus covering fraction. To make the contamination
from the host galaxy least, here we adopt the MIR photometric results 
based on high spatial-resolution observations by \citet{Asmus2014}.
Compared with the diagnostics using the $L_{\rm [O\ IV]}/L_{\rm X}$ ratio, this
method has advantages that (1) X-ray spectra are not required 
(applicable even for heavily Compton thick AGNs), and that (2) it is
little affected by the UV-to-X-ray SED. In Section 2, to confirm the
validity of our method, we apply this diagnostics to a \textit{Swift}/BAT
selected AGN sample and compare the results with those of the X-ray
scattering fraction. In Section 3, we apply it to a sample of local
U/LIRGs having AGNs, and discuss the fraction of buried AGNs in
different merger stages. In Section 4, we summarize our work. Following
\citet{Asmus2014}, we adopt the cosmological parameters ($H_{\rm 0}$,
$\Omega_{m}$, $\Omega_{\rm \lambda}$) $=$ (67.3, 0.315, 0.685) 
\citep{Planck2014} throughout the paper.

\begin{deluxetable*}{lcrrrcccccc}
\tablewidth{\textwidth}
\tablecaption{X-Ray and MIR Emission Properties of Absorbed AGNs in the \textit{Swift}/BAT 9-Month Catalog \label{tab1-swift}}
\tablehead{
\colhead{Object}      &
\colhead{$z$} &
\colhead{$D_{\rm L}$} &
\colhead{$f_{\rm scat}$}          &
\colhead{log$N_{\rm H}$} &
\colhead{log$L_{\rm X}$}   &
\colhead{Ref. 1} &
\colhead{log$L^{\rm (nuc)}_{\rm 12\,\mu m}$} &
\colhead{log$L_{\rm [O\,IV]}$} &
\colhead{Ref. 2}\\
\ \ \ \ \ \ \ \ (1)&(2)&(3)\ \ &(4)\ \ \ \ \ \ &(5)\ \ \ \ \ \ \ &(6)&(7)&(8)&(9)&(10)
}
\startdata
NGC 235A      & 0.02223  &  96.2 & 0.51$^{+0.23}_{-0.18}$ & 23.81$^{+0.05}_{-0.04}$ & 43.10 & 1 & 43.31 $\pm$ 0.20 & 41.28 $\pm$ 0.01  & 5 \\
ESO 297--18   & 0.02523  & 111.0 & 0.26$^{+0.15}_{-0.14}$ & 23.80 $\pm$ 0.02        & 43.50 & 1 & 43.06 $\pm$ 0.07 & 40.69             & 6 \\
NGC 788       & 0.01360  &  57.2 & 0.71$^{+0.19}_{-0.14}$ & 23.87 $\pm$ 0.02        & 43.00 & 1 & 43.12 $\pm$ 0.05 & 40.98 $\pm$ 0.01  & 7 \\
NGC 2110      & 0.00779  &  35.9 & 0.28 $\pm$ 0.04        & 22.37 $\pm$ 0.01        & 43.30 & 1 & 43.09 $\pm$ 0.06 & 40.85 $\pm$ 0.03  & 7 \\
MCG--1--24--12& 0.01964  &  93.8 & 0.50$^{+0.30}_{-0.20}$ & 22.81$^{+0.05}_{-0.03}$ & 43.24 & 2 & 43.46 $\pm$ 0.04 & 41.03 $\pm$ 0.03  & 7 \\
MCG--5--23--16& 0.00850  &  42.8 & 0.47$^{+0.04}_{-0.03}$ & 22.20 $\pm$ 0.01        & 43.30 & 1 & 43.26 $\pm$ 0.04 & 40.79 $\pm$ 0.12  & 7 \\
NGC 3081      & 0.00798  &  40.9 & 0.52$^{+0.13}_{-0.10}$ & 23.92 $\pm$ 0.02        & 42.30 & 1 & 42.91 $\pm$ 0.11 & 41.38 $\pm$ 0.03  & 7 \\
NGC 3281      & 0.01067  &  52.8 & 1.90$^{+2.50}_{-1.20}$ & 23.94 $\pm$ 0.08        & 42.69 & 3 & 43.61 $\pm$ 0.05 & 41.77 $\pm$ 0.03  & 7 \\
NGC 4507      & 0.01180  &  57.5 & 0.31$^{+0.05}_{-0.04}$ & 23.43$^{+0.08}_{-0.06}$ & 43.10 & 1 & 43.79 $\pm$ 0.04 & 41.16 $\pm$ 0.05  & 7 \\
ESO 506--27   & 0.02502  & 119.0 & 0.34$^{+0.10}_{-0.08}$ & 23.92$^{+0.03}_{-0.02}$ & 43.30 & 1 & 43.88 $\pm$ 0.04 & 40.82             & 6 \\
NGC 4992      & 0.02514  & 119.0 & 0.00$^{+0.17}$         & 23.78$^{+0.03}_{-0.02}$ & 43.30 & 1 & 43.53 $\pm$ 0.10 & 40.27             & 6 \\
NGC 5252      & 0.02297  & 109.0 & 0.39$^{+0.24}_{-0.32}$ & 22.32 $\pm$ 0.11        & 43.10 & 1 & 43.39 $\pm$ 0.04 & 41.09             & 6 \\
NGC 5506      & 0.00618  &  31.6 & 1.09$^{+0.04}_{-0.05}$ & 22.49 $\pm$ 0.01        & 43.00 & 1 & 43.41 $\pm$ 0.03 & 41.48 $\pm$ 0.01  & 7 \\
ESO 103--35   & 0.01329  &  59.5 & 0.10$^{+0.03}_{-0.02}$ & 23.31 $\pm$ 0.01        & 43.50 & 1 & 43.71 $\pm$ 0.23 & 41.16 $\pm$ 0.01  & 7 \\
IC 5063       & 0.01135  &  49.1 & 0.90 $\pm$ 0.10        & 23.40 $\pm$ 0.01        & 43.08 & 4 & 43.77 $\pm$ 0.03 & 41.53 $\pm$ 0.04  & 7 \\
NGC 7172      & 0.00868  &  34.8 & 0.12 $\pm$ 0.03        & 22.95 $\pm$ 0.01        & 43.00 & 1 & 42.83 $\pm$ 0.04 & 40.79 $\pm$ 0.04  & 7
\enddata
\tablecomments{Columns:
(1) Object name;
(2) redshift from NASA/IPAC Extragalaxtic Database (NED);
(3) luminosity distance in Mpc from \citet{Asmus2014};
(4) scattering fraction in percent;
(5) X-ray absorption hydrogen column density in cm$^{-2}$;
(6) absorption corrected 2--10~keV luminosity in erg s$^{-1}$;
(7) references of columns (4)--(6);
(8) nuclear subarcsecond-scale monochromatic luminosities at rest-frame 12 $\mu$m \citep{Asmus2014};
(9)--(10) [\ion{O}{4}] 25.89 $\mu$m luminosity in erg s$^{-1}$ and its reference;
list of references:
1: \citet{Kawamuro2016a}; 
2: \citet{Ricci2017d};
3: \citet{Winter2009};
4: \citet{Tazaki2011};
5: \citet{Inami2013};
6: \citet{Weedman2012};
7: \citet{Weaver2010}.
}
\end{deluxetable*}

\section{$L_{\rm [O\ IV]}$/$L^{\rm (\MakeLowercase{nuc})}_{\rm 12\,\mu \MakeLowercase{m}}$ Ratio as Diagnostics of Torus Structure}
\label{sec2}

To justify that the ratio of the [\ion{O}{4}] to nuclear 12 $\mu$m luminosities
is a good indicator for the opening angle of the torus, we first
investigate the correlation between this ratio and the X-ray scattering
fraction ($f_{\rm scat}$), using AGNs in the 
\textit{Swift}/BAT 9-month catalog \citep{Tueller2008}. 
Limiting the sample to non-blazar AGNs at Galactic latitudes of
$|b|>15^\circ$, we basically 
refer to the results of X-ray spectral analysis compiled by
\citet{Ueda2015}, but update them with more recent work by \citet{Kawamuro2016a}
based on \textit{Suzaku} observations whenever
available. The $f_{\rm scat}$ value is defined as the ratio of the
unabsorbed fluxes at 1 keV between the scattered and transmitted
components.
We select only Compton-thin absorbed AGNs (log$N_{\rm
H}$/cm$^{-2} =$ 22--24)\footnote{Because the scattered components are
measurable only in absorbed AGNs. We also exclude Compton-thick AGNs
(log$N_{\rm H}$/cm$^{-2} >$ 24) to avoid possible systematic
uncertainties in the intrinsic luminosities (and hence in $f_{\rm
scat}$), which largely depend on spectral models adopted 
\citep[see e.g.,][]{Murphy2009,Tanimoto2018}.} whose scattering
fractions are less than 2\%\footnote{Because apparently large $f_{\rm scat}$
values are likely to be caused by partial absorbers in the line of sight
\citep[see][]{Ueda2015}.}. 
To make sure that these $f_{\rm scat}$ values are reliable, 
we check if the contamination from X-ray binaries 
in the host galaxy significantly affects 
the flux of the unabsorbed power-law component.
Using the total infrared luminosity 
(8--1000~$\mu$m)\footnote{We 
convert 40--120~$\mu$m luminosity to 8--1000~$\mu$m band 
using a conversion factor of 1.9 \citep{Lutz2018}.} estimated by
\citet{Lutz2018} and the relation between total infrared 
and X-ray luminosity by \citet{Mineo2012},
we find that the estimated contribution from X-ray binaries
is almost ignorable (the ratio of the total luminosity of X-ray binaries
to the intrinsic luminosity of the AGN is 
$\lesssim$ 0.1\% in the 0.5--2~keV band)
except for NGC~4388 and NGC~4138 
(0.3\% and 0.8\%, respectively), which we exclude from our analysis.
Thus, our final sample consists of 16 objects. 

Then, the [\ion{O}{4}] 25.89 $\mu$m luminosities for these AGNs are taken from
\citet{Weaver2010} and \citet{Weedman2012}, which are based on the
MIR spectra observed with the Infrared Spectrograph (IRS; \citealt{Houck2004})
on board the \textit{Spitzer} observatory. We refer to \citet{Asmus2014}
for the 12 $\mu$m photometry at 
a~$\lesssim$~0.4 arcsec scale, corresponding to
$\lesssim$~300~pc at z~$\lesssim$~0.03,
observed with the Very-Large-Telescope mounted Spectrometer and Imager
for the Mid-infrared (VISIR; \citealt{Lagage2004}), the
\textit{Gemini}/Michelle \citep{Glasse1997}, and the
\textit{Subaru}/Cooled Mid-Infrared Camera and Spectrometer (COMICS;
\citealt{Kataza2000}). 
All the 16 \textit{Swift}/BAT AGNs selected above have the measurements of 
both [\ion{O}{4}] 25.89 $\mu$m and nuclear 12 $\mu$m luminosities, as summarized in Table~1. 

\begin{figure*}
    \epsscale{1.14}
    \plottwo{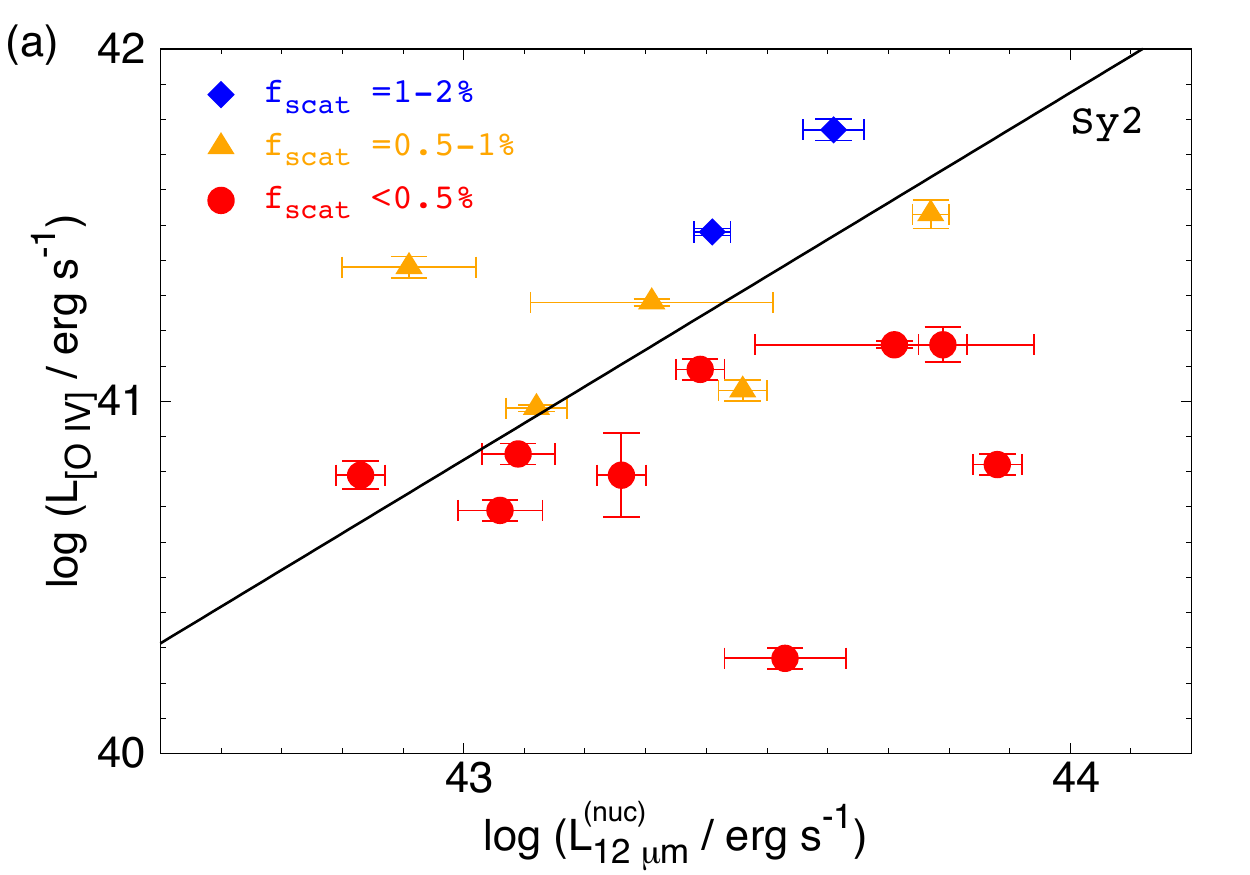}{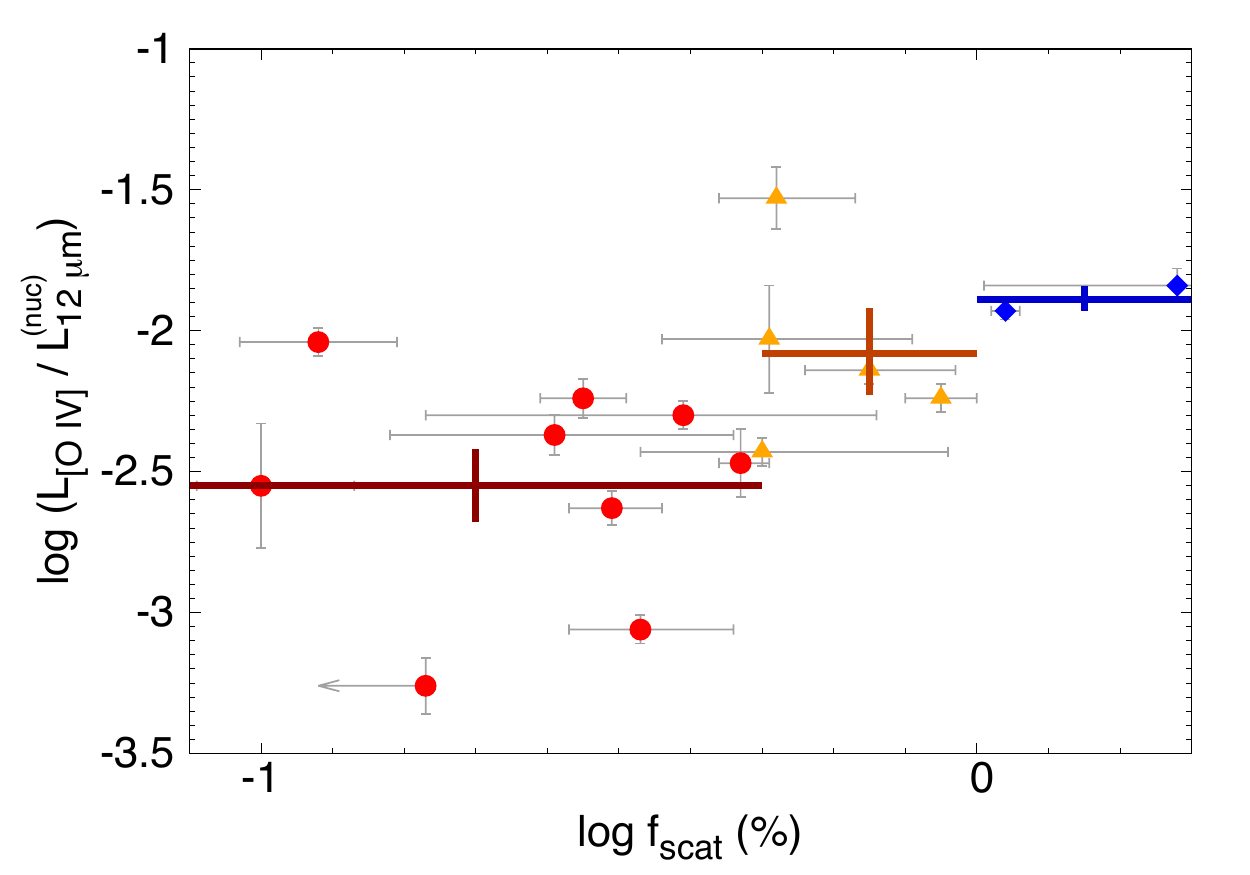}
    \caption{(a) [\ion{O}{4}] 25.89 $\mu$m luminosity vs. nuclear 12 $\mu$m
    luminosity for our sample in the \textit{Swift}/BAT 9-month catalog.
    Circles represent Compton-thin AGNs in Table 1.
    Red circles, orange triangles, and blue diamonds mark AGNs whose scattering fractions are $<$0.5\%, 0.5--1\%, and 1--2\%, respectively.
    The solid line shows the averaged relation for Seyfert 2s obtained by \citet{Yang2015}. \ 
    (b) The ratio of [\ion{O}{4}] to nuclear 12 $\mu$m luminosities vs. scattering fraction.
    Bold crosses represent the mean values (center) and 
    standard errors (half length of the vertical bar) in the three $f_{\rm scat}$ bins.
    \\}
    \label{fig1-oiv-12um}
\end{figure*}

Figure 1(a) plots the luminosity correlations between the [\ion{O}{4}] 25.89
$\mu$m line and nuclear 12 $\mu$m continuum for the AGNs in Table 1. As
noticed, the low X-ray-scattering AGNs ($f_{\rm scat} <$ 0.5\%) show
smaller [\ion{O}{4}] luminosities relative to the nuclear 12 $\mu$m ones than
the average of Seyfert 2s obtained by \citet{Yang2015}.
Figure 1(b) plots the $L_{\rm [O\ IV]}/L^{\rm (nuc)}_{\rm 12\,\mu m}$
ratio as a function of 
the X-ray scattering fraction, together with 
the mean values and 
standard errors in three $f_{\rm scat}$ bins.
We find that the $L_{\rm [O\ IV]}/L^{\rm (nuc)}_{\rm 12\,\mu m}$ ratios tend to 
be lower for AGNs with lower $f_{\rm scat}$ values;
the averaged $L_{\rm [O\ IV]}/L^{\rm (nuc)}_{\rm 12\,\mu m}$ ratio of the low X-ray-scattering
AGNs (log$L_{\rm [O\ IV]}/L^{\rm (nuc)}_{\rm 12\,\mu m}$ = $-2.6 \pm 0.1$) is 
smaller than those of the others ($\sim-2.0$).
A Kolmogorov-Smirnov (KS) test 
indicates that the difference in the 
$L_{\rm [O\ IV]}/L^{\rm (nuc)}_{\rm 12\,\mu m}$ distribution
between the low X-ray-scattering AGNs and the others
is significant at a $>$99\% confidence level.

\citet{Yang2015} suggest that both X-ray and continuum MIR
emission in AGNs may be mildly anisotropic by assuming that
the [\ion{O}{4}] 25.89 $\mu$m line is an isotropic luminosity indicator;
AGNs viewed at higher inclination angles tend to have smaller
X-ray and MIR luminosities. This would lead to
an apparent correlation between $f_{\rm scat}$ ($\propto 1/L_{\rm X}$) and
$L_{\rm [O\ IV]}/L^{\rm (nuc)}_{\rm 12\,\mu m}$. However, since our
sample consists of only obscured AGNs, such effect, if present, 
would be limited. In fact, 
\citet{Liu2014} report that the
correlation between log$L_{\rm [O\ IV]}$ and log$L_{\rm X}$ are statistically indistinguishable
between Seyfert 2s with $10^{23} < N_{\rm H} < 10^{24}$ cm$^{−2}$ and
those with $N_{\rm H} < 10^{23}$ cm$^{−2}$, which are likely to have
relatively high and low inclination angles among obscured AGNs, respectively.
According to the clumpy torus
model by \citet{Stalevski2016}, the difference in the MIR
luminosity is within a factor of 1.5 for an inclination angle range of
60$^\circ$--90$^\circ$. This is smaller than the observed 
difference of the mean $L_{\rm [O\
IV]}/L^{\rm (nuc)}_{\rm 12\,\mu m}$ ratio between the 
$f_{\rm scat} <$ 0.5\% and $f_{\rm scat} >$ 0.5\% samples 
(a factor of $\sim$4).
Hence, the observed 
$f_{\rm scat}$--$L_{\rm [O\ IV]}/L^{\rm (nuc)}_{\rm 12\,\mu m}$ correlation 
cannot be accounted for by 
the possible anisotropy effects.
These results support that the $L_{\rm [O\ IV]}/L^{\rm (nuc)}_{\rm 12\,\mu m}$ ratio 
is indeed a good estimator of the torus opening angle and hence can be used 
to identify whether an AGN is buried.

\section{Application to U/LIRG}
\label{sec3}
We apply our $L_{\rm [O\ IV]}/L^{\rm (\MakeLowercase{nuc})}_{\rm 12\,\mu m}$ method to local U/LIRGs in the
Great Observatories All-sky LIRG Survey (GOALS; \citealt{Armus2009}). The
GOALS targets 202 U/LIRGs in the local universe contained in the IRAS
Revised Bright Galaxy Sample (RBGS; \citealt{Sanders2003}), which is a
complete flux limited sample of 629 extragalactic objects having $f_{\rm
\nu}$(60 $\mu$m) $>$~5.24~Jy at Galactic latitudes $|b| > 5^{\circ}$.
Using high-resolution spectra obtained with \textit{Spitzer}/IRS,
\citet{Inami2013} estimate the [\ion{O}{4}] 25.89 $\mu$m
fluxes of all GOALS samples.
The nuclear 12 $\mu$m fluxes are
obtained by \citet{Asmus2014} for 23 U/LIRGs in the GOALS sample,
which constitute the sample for our study.
We summarize their MIR properties in Table 2\footnote{We exclude NGC
6240 and NGC 1275 from our sample by the following reasons. NGC 6240
hosts dual AGNs as revealed by Chandra \citep{Komossa2003} but the [\ion{O}{4}] 
fluxes are not separated.  NGC 1275 is a radio galaxy located in the
Perseus cluster, and its 12 $\mu$m flux may be contaminated with the
emission from the jet and the cluster \citep[see][]{Hitomi2018}. }.
Merger stages of the GOALS sample are specified by \citet{Stierwalt2013},
via visual inspection of the IRAC 3.6 $\mu$m images 
and/or higher resolution images in the
literature. Each U/LIRG is assigned one of the following five
designations: non-mergers (no sign of merger activity or massive
neighbors: ``N''), pre-mergers (galaxy pairs prior to a first encounter:
``A''), early stage mergers (post-first-encounter with galaxy disks
still symmetric and intact but with signs of tidal tails: ``B''),
mid-stage mergers (showing amorphous disks, tidal tails, and other signs
of merger activity: ``C''), or late stage mergers (two nuclei in a
common envelope: ``D'').

\begin{deluxetable*}{lcrclccccccc}
\tablewidth{\textwidth}
\tablecaption{MIR Emission Properties of AGNs in the GOALS sample \label{tab2-goals}}
\tablehead{
\colhead{Object}   &
\colhead{$z$}        &
\colhead{$D_{\rm L}$}  &
\colhead{log$L_{\rm IR}/L_{\odot}$}        &
\colhead{M}    &
\colhead{Type}&
\colhead{$\langle\alpha^{\rm (MIR)}_{\rm AGN}\rangle$} &
\colhead{log$N_{\rm H}$}&
\colhead{Ref}&
\colhead{log$L^{\rm (nuc)}_{\rm 12\,\mu m}$}&
\colhead{log$L_{\rm [O\,IV]}$} \\
\ \ \ \ \ \ \ \ \ \ \ \ (1)&(2)&(3)\ \ &(4)&(5)&(6)&(7)&(8)&(9)&(10)&(11)
}
\startdata
NGC 34           & 0.01962 &  83.5 & 11.49$^{\ \,}$    & \,D       &     Sy2 & 0.20 $\pm$ 0.11 &    23.7 &       1 & 43.08 $\pm$ 0.05 & $<$ 40.67\ \  \\
NGC 235A         & 0.02223 &  96.2 & 11.44$^{a}$       & \,B       &     Sy2 & 0.59 $\pm$ 0.11 &    23.5 &       2 & 43.31 $\pm$ 0.20 & 41.28 $\pm$ 0.01 \\
NGC 1068         & 0.00379 &  14.4 & 11.40$^{\ \,}$    & \,N$^{b}$ & Sy1.8/2 & 1.00            &$>$ 24.0 &       3 & 43.80 $\pm$ 0.15 & 41.70 $\pm$ 0.28 \\
NGC 1365         & 0.00546 &  17.9 & 11.00$^{\ \,}$    & \,N       &   Sy1.8 & 0.51 $\pm$ 0.14 &    24.0 &       4 & 42.54 $\pm$ 0.04 & 40.72 $\pm$ 0.03 \\
ESO 420--13      & 0.01191 &  52.7 & 11.07$^{\ \,}$    & \,N       &      Cp & 0.42 $\pm$ 0.12 & \nodata & \nodata & 43.20 $\pm$ 0.09 & 41.25 $\pm$ 0.01 \\
IRAS 05189--2524 & 0.04256 & 196.0 & 12.16$^{\ \,}$    & \,D       &     Sy2 & 0.67 $\pm$ 0.10 &    23.1 &       1 & 44.87 $\pm$ 0.17 & 42.08 $\pm$ 0.02 \\
NGC 2623         & 0.01851 &  87.3 & 11.60$^{\ \,}$    & \,D       &      Cp & 0.26 $\pm$ 0.10 &$>$ 24.0 &       5 & 43.61 $\pm$ 0.26 & 40.92 $\pm$ 0.02 \\
IRAS F08572+3915 & 0.05835 & 275.0 & 12.16$^{\ \,}$    & \,D       &  \,\ Cp*& 0.46 $\pm$ 0.17 & \nodata & \nodata & 45.13 $\pm$ 0.09 & $<$ 41.41 \\
UGC 5101         & 0.03937 & 182.0 & 12.01$^{\ \,}$    & \,D       &      Cp & 0.41 $\pm$ 0.16 &    24.1 &       1 & 44.35 $\pm$ 0.08 & 41.46 $\pm$ 0.01 \\
NGC 3690W        & 0.01022 &  48.2 & 11.93$^{a}$       & \,C       &      Cp & 0.46 $\pm$ 0.15 &    24.6 &       6 & 43.73 $\pm$ 0.28 & 40.90 $\pm$ 0.04 \\
NGC 3690E        & 0.01041 &  49.1 & 11.93$^{a}$       & \,C       &      Cp & 0.17 $\pm$ 0.04 &    22.1 &       7 & 43.30 $\pm$ 0.24 & 40.67 $\pm$ 0.15 \\
IC 883           & 0.02330 & 109.0 & 11.73$^{\ \,}$    & \,D       &  \,\ Cp$^{c}$& 0.13 $\pm$ 0.08 &    21.3 &       8 & $<$ 43.90 \ \ \ \ & 40.99 $\pm$ 0.01 \\
MCG--03--34--064 & 0.01654 &  79.3 & 11.28$^{\ \,}$    & \,A       & Sy1.8/2 & 0.85 $\pm$ 0.07 &    23.7 &       1 & 44.00 $\pm$ 0.05 & 41.90 $\pm$ 0.02 \\
NGC 5135         & 0.01369 &  66.0 & 11.30$^{\ \,}$    & \,N       &     Sy2 & 0.36 $\pm$ 0.08 &    24.8 &       9 & 43.24 $\pm$ 0.08 & 41.61 $\pm$ 0.01 \\
IC 4518W         & 0.01626 &  76.1 & 11.23$^{\ \,}$    & \,B$^{d}$ &     Sy2 & \nodata         &    23.4 &       1 & 43.54 $\pm$ 0.07 & 41.77$^{d}$ \\
IRAS F15250+3608 & 0.05516 & 258.0 & 12.08$^{\ \,}$    & \,D       & \,\ Cp:*& 0.24 $\pm$ 0.21 & \nodata & \nodata & 44.64 $\pm$ 0.04 & $<$ 41.52 \\
ESO 286--19      & 0.04300 & 195.0 & 12.06$^{\ \,}$    & \,D       & \,\ Cp:*& 0.39 $\pm$ 0.15 & \nodata & \nodata & 44.67 $\pm$ 0.04 & $<$ 41.28 \\
NGC 7130         & 0.01615 &  68.9 & 11.42$^{\ \,}$    & \,N       &      Cp & 0.39 $\pm$ 0.12 &    24.6 &       1 & 43.17 $\pm$ 0.09 & 41.07 $\pm$ 0.01 \\
ESO 602--25      & 0.02504 & 109.0 & 11.34$^{\ \,}$    & \,N       & \,\ Cp:$^{c}$& 0.31 $\pm$ 0.11 & \nodata & \nodata & 43.08 $\pm$ 0.17 & 40.90 $\pm$ 0.02 \\
NGC 7469         & 0.01632 &  67.9 & 11.65$^{\ \,}$    & \,A       & Sy1/1.5 & 0.44 $\pm$ 0.14 &    19.8 &       1 & 43.83 $\pm$ 0.05 & 41.43 $\pm$ 0.02 \\
NGC 7592W         & 0.02462 & 105.0 & 11.40$^{a}$      & \,B       &      Cp & 0.40 $\pm$ 0.13 & \nodata & \nodata & 43.65 $\pm$ 0.16 & 40.92 $\pm$ 0.01 \\
NGC 7674         & 0.02892 & 126.0 & 11.56$^{\ \,}$    & \,A       &     Sy2 & 0.80 $\pm$ 0.07 &$>$ 24.5 &       1 & 44.26 $\pm$ 0.06 & 41.94 $\pm$ 0.01 \\
NGC 7679         & 0.01714 &  71.7 & 11.11$^{\ \,}$    & \,A       &      Cp & 0.27 $\pm$ 0.09 &$<$ 20.3 &       1 & 42.74 $\pm$ 0.12 & 41.32 $\pm$ 0.01
\enddata

\tablecomments{Columns:
(1) Object name;
(2) redshift from NED;
(3) luminosity distance in Mpc;
(4) total infrared luminosity in units of $L_{\odot}$;
(5) merger stage (N = non-merger, A = pre-merger, B = early stage merger, C = mid-stage merger, and D = late stage merger) as classified from the \textit{Hubble Space Telescope} (\textit{HST}) and IRAC 3.6 $\mu$m imaging (see \citealt{Stierwalt2013});
(6) optical AGN classification;
Cp means that the object has been classified as AGN/starburst composites, and
the suffix `:' means that the classification is uncertain;
Objects marked with an * mark uncertain AGNs;
(7) AGN fractional contribution to the total MIR luminosity based on five \textit{Spitzer}/IRS diagnostics from \citet{Diaz-Santos2017};
(8)--(9) X-ray absorption hydrogen column-density in cm$^{-2}$ and its reference;
(10) nuclear subarcsecond-scale monochromatic luminosities at rest-frame 12 $\mu$m;
(11) [\ion{O}{4}] 25.89 $\mu$m luminosity in erg s$^{-1}$.
Columns (3), (6), and (10) are taken from \citet{Asmus2014}.
Column (4) and (11) are taken from \citet{Armus2009} and \citet{Inami2013}, respectively.
List of references:
1: \citet{Ricci2017b};
2: \citet{Winter2009};
3: \citet{Ricci2014};
4: \citet{Rivers2015};
5: \citet{Evans2008};
6: \citet{Ptak2015};
7: \citet{Zezas2003};
8: \citet{Romero-Canizales2017};
9: Yamada et al. in prep. (using the \textit{NuSTAR} observations.)
}
\tablenotetext{a}{Including the emission from the companion.}
\tablenotetext{b}{\citet{Tanaka2017} suggest that NGC 1068 may be tidally induced structures of a past minor merger.}
\tablenotetext{c}{Although these are classified as uncertain AGNs, the [\ion{Ne}{5}] 14.32 $\mu$m lines are detected \citep{Inami2013}.}
\tablenotetext{d}{The merger stage and [\ion{O}{4}] 25.89 $\mu$m luminosity are following \citet{Ricci2017b} and \citet{Yang2015}, respectively.}
\end{deluxetable*}

Contribution from star formation to the MIR continuum may be significant
in a U/LIRG, even if we refer to the nuclear flux.
In fact, many objects in Table 2 are classified as AGN/starburst composites.
\citet{Diaz-Santos2017} estimate the fractional AGN contribution to
the MIR luminosity for the GOALS sample with
\textit{Spitzer}/IRS observations
\footnote{The projected aperture sizes are 
3".7 $\times$ 12" at 9.8 $\mu$m
in SL module, 10".6 $\times$ 35" at 26 $\mu$m in LL, 4".7 $\times$ 15".5 at 14.8 $\mu$m in
SH, and 11".1 $\times$ 36".6 at 28 $\mu$m in LH
\citep{Diaz-Santos2017}.}, 
by employing up
to five diagnostics: the line flux ratios of
[\ion{Ne}{5}]$_{14.3}$/[\ion{Ne}{2}]$_{12.8}$ and [\ion{O}{4}]$_{25.9}$/[\ion{Ne}{2}]$_{12.8}$,
the equivalent width of the 6.2 $\mu$m Polycyclic Aromatic Hydrocarbons
(PAH), the $S_{30}$/$S_{15}$ dust continuum slope, and the Laurent
diagram \citep{Laurent2000}. The averaged estimate of the MIR AGN fraction for each galaxy,
$\langle\alpha^{\rm (MIR)}_{\rm AGN}\rangle$, 
is listed in Table~2. 
To check their systematic uncertainties,
we also calculate the mean MIR AGN fractions by excluding the 
[\ion{Ne}{5}]/[\ion{Ne}{2}] and [\ion{O}{4}]/[\ion{Ne}{2}] diagnostics,
because the [\ion{Ne}{5}] and [\ion{O}{4}] lines 
may not be good AGN-power indicators in buried AGNs.
We find that then the estimates become larger 
than those obtained with the five diagnostics by a factor of 1.1--1.7.
This implies that the AGN fractions in Table~2 may be 
underestimated; nevertheless, in the following, we adopt these numbers 
for conservative discussions.
It is confirmed that generally the composites have 
smaller $\langle\alpha^{\rm (MIR)}_{\rm AGN}\rangle$ than
Seyfert~1.8/2s. In our paper, we distinguish U/LIRGs with 
$\langle\alpha^{\rm (MIR)}_{\rm AGN}\rangle <$ 1/3 
as starburst-dominant objects, and the rest as AGN-important ones.
\citet{Asmus2015} find that AGN-important objects (e.g.,
NGC 3690W, NGC 7130) follow the same MIR-X-ray luminosity correlation
as for normal Seyferts, while the starburst-dominant ones (e.g., NGC
3690E) do not.

\begin{figure*}
    \epsscale{1.14}
    \plottwo{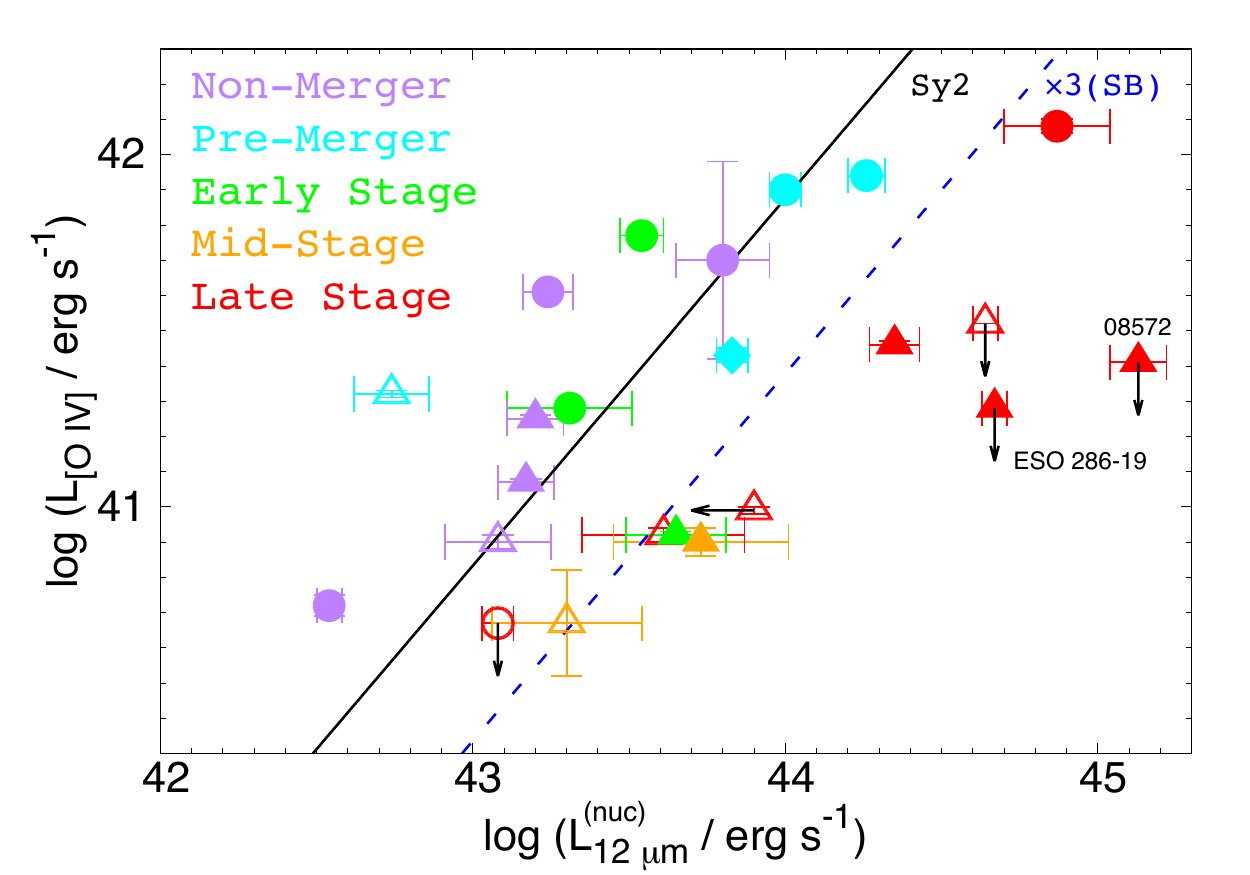}{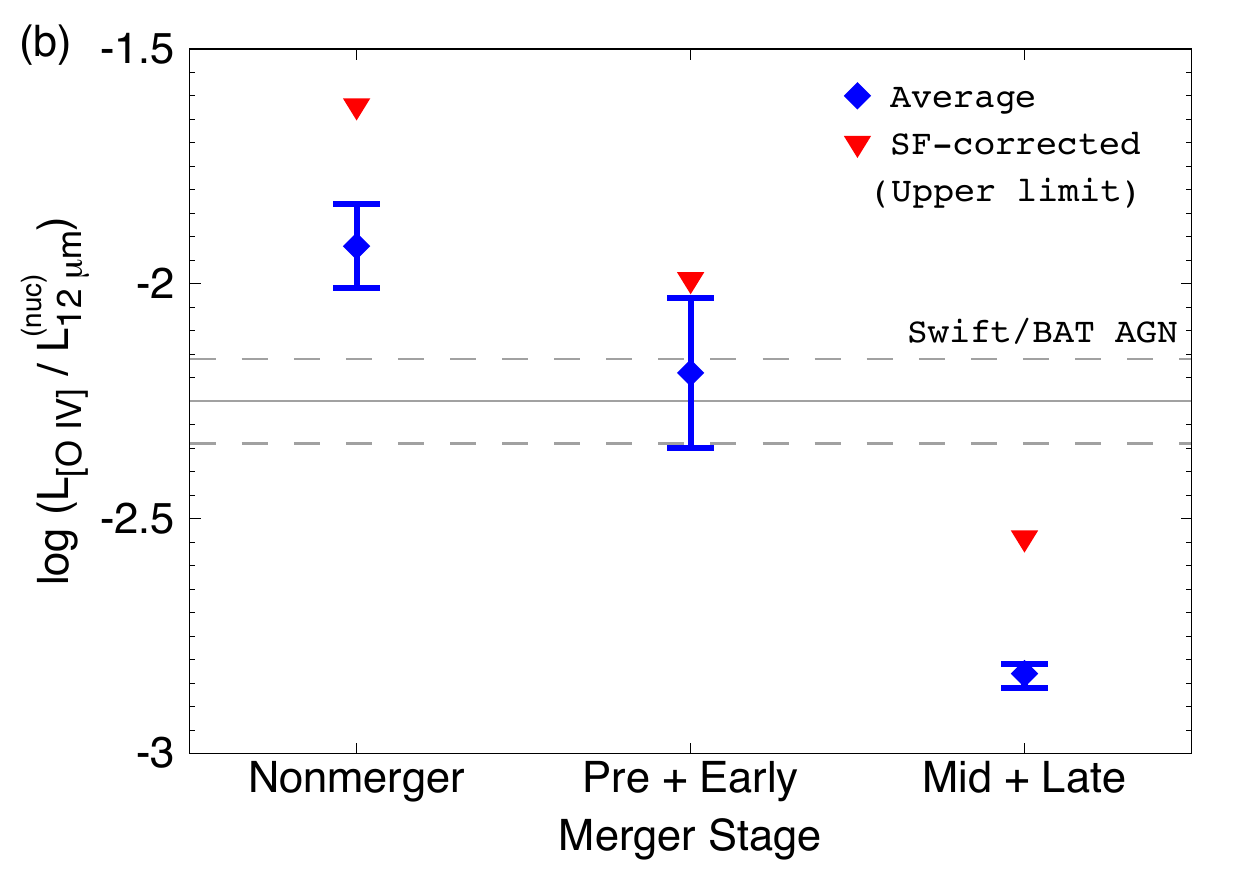}
    \caption{(a) [\ion{O}{4}] 25.89 $\mu$m luminosity vs. nuclear 12 $\mu$m luminosity for the GOALS samples in Table 2.
    These plots are color coded by the merger stages determined by \citet{Stierwalt2013}.
    Diamond, circles, and triangles mark the Seyfert 1/1.5, Seyfert 1.8/2s, and AGN/starburst composites (Cp/Cp:), respectively.
    Empty symbols are the starburst-dominated objects whose MIR AGN fraction is $<$1/3.
    Arrows mark upper limits in the case of MIR non-detection.
    Black-solid and blue-dashed lines show the averaged relation for Seyfert 2s obtained by \citet{Yang2015} and that corrected for contribution of starburst in the MIR luminosity by a factor of 3.\
    (b) The ratio of [\ion{O}{4}] to nuclear 12 $\mu$m luminosities vs. merger stage.
    The gray solid and dashed lines correspond to the average and standard deviation, respectively, for the \textit{Swift}/BAT sample in Table 1.
    Blue diamonds represent the mean values (center) and 
    standard errors (half length of the
    vertical bar) in the three merger stages.
    Red triangles represent the results where
    the star formation contributions are subtracted from 
    the nuclear 12 $\mu$m luminosities by using the 
    MIR AGN fractions in Table~2. They correspond to upper limits (see the text).
    \\}
    \label{fig3-LIRGs}
\end{figure*}

In Figure~2(a), we present the correlation between 
[\ion{O}{4}] and nuclear 12 $\mu$m luminosities for the U/LIRGs in Table~2.
The merger stages are distinguished with colors. 
We find that the 
non-mergers and pre-/early-stage mergers, most of which 
are Seyfert 1.8/2s or AGN-important composites (Table~2), 
follow the same correlation as for typical Seyfert 2s 
(the black solid line; \citealt{Yang2015}). By contrast, the 
mid- and late-stage mergers, most of which are composites, 
show smaller $L_{\rm [O\ IV]}/L^{\rm (nuc)}_{\rm 12\ \mu m}$ ratios
than typical Seyfert 2s. 
We need to keep in mind, however, that 
their 12 $\mu$m luminosities may 
be largely contaminated by the starburst activities.
Hence, we hereafter exclude the starburst-dominant objects (empty
symbols) from our discussion. 

Focusing on the AGN-important objects (filled symbols), we plot the
expected $L_{\rm [O\ IV]}/L^{\rm (nuc)}_{\rm 12\ \mu m}$ correlation corrected for the maximum
contribution from starburst (i.e., by a factor of 3)\footnote{Here 
we assume that the 12 $\mu$m luminosity 
is proportional to the MIR one.}
by the dashed line
in Figure 2(a). This line should be taken as 
lower limits of $L_{\rm [O\ IV]}/L^{\rm (nuc)}_{\rm 12\ \mu m}$, 
because (1) it assumes the {\it minimum} MIR AGN fraction (1/3), 
(2) the MIR AGN fractions in Table~2 may be underestimated (see above),
and (3) we utilize the {\it nuclear} MIR flux, instead of 
those from the whole galaxy as used by \citet{Diaz-Santos2017}.
Among the five AGN-important objects in mid-/late-stage
mergers, we regard IRAS F05189--2524, UGC~5101, and NGC~3690W as certain
AGNs, whose hard X-ray transmitted components are detected with \textit{NuSTAR}
\citep[][respectively]{Teng2015,Oda2017,Ptak2015}. The rest two objects, IRAS 08572+3915 and ESO 286--19, are
classified as ``uncertain AGNs'' in \citet{Asmus2014}. Nevertheless, 
infrared observations suggest that they have ``buried'' AGNs (2.5--5 $\mu$m, \citealt{Imanishi2008}; 5--8~$\mu$m, \citealt{Nardini2010}). 
As noticed from Figure~2(a), all the AGN-important mid-/late-stage
mergers have lower $L_{\rm [O\ IV]}/L^{\rm (nuc)}_{\rm 12\ \mu m}$ ratios than the correlation of
typical Seyfert galaxies corrected for the maximum starburst
contribution, indicating that they are buried by the tori with large 
covering fractions.

Figure 2(b) represents the mean values and 
standard errors of the
log$L_{\rm [O\ IV]}/L^{\rm (nuc)}_{\rm 12\ \mu m}$ ratio
in three different merger stages. 
For robust discussion, here we have excluded the
Seyfert~1/1.5 (NGC 7469), starburst-dominated objects, and 
uncertain AGNs.
In addition, we also plot 
the results where the star 
formation contributions are subtracted 
from the nuclear 12 $\mu$m luminosities
by using the MIR AGN fractions in Table~2 (red triangles).
As mentioned above, these MIR AGN fractions are likely to be underestimated
and hence the resultant log$L_{\rm [O\ IV]}/L^{\rm
(nuc)}_{\rm 12\ \mu m}$ values should be taken as their upper limits.
A trend is seen that the [\ion{O}{4}]-to-12~$\mu$m ratio decreases 
with merger stage.
The non-mergers and pre-/early-stage mergers show
mean values similar to typical Seyfert~2s 
(log$L_{\rm [O\ IV]}/L^{\rm (nuc)}_{\rm 12\ \mu m}$ $\approx -2.1$; \citealt{Yang2015}) and to 
the \textit{Swift}/BAT sample ($-2.3 \pm 0.1$; Table~1), respectively.
By contrast, 
the mid-/late-stage mergers show a 
much smaller value ($\approx
-2.8$); this is true
even if we refer to the upper limit corrected for star formation contributions 
($-2.5$).
A KS test indicates that 
the log$L_{\rm [O\ IV]}/L^{\rm (nuc)}_{\rm 12\ \mu m}$ distribution 
is different at a $>$99\% confidence level
between the mid-to-late stage mergers and the others.
Thus, our results support the scenario that the fraction of buried AGNs
in U/LIRGs increases as the galaxy-galaxy interaction becomes more
significant.

Our conclusion is well in line with the hard X-ray results \citep{Ricci2017b},
who show that the torus covering fraction is very high
$95^{+4}_{-8}$\% in late stage mergers on the basis of statistical
argument. Our method has a potential to be applied to individual U/LIRGs
for which MIR spectroscopy is available but the current sensitivities of
hard X-ray observations are insufficient. We note that instead of using
the spatially-resolved nuclear flux, spectral decomposition of the IR
SED should be also useful in estimating the AGN contribution in the MIR
fluxes \citep{Ichikawa2019}. This is a subject of future work.

\section{Conclusion}
\label{sec4}
For identifying whether an AGN is deeply buried, 
we propose new diagnostic that utilizes the
ratio between [\ion{O}{4}] 25.89 $\mu$m and nuclear 12 $\mu$m luminosities.
First, to confirm the validity of
this diagnostics, we investigate the relation between this ratio and the
X-ray scattering fraction, using a sample of 16 \textit{Swift}/BAT AGNs observed
with both \textit{Spitzer}/IRS and ground-based high angular-resolution
MIR cameras. We find that low X-ray-scattering AGNs with $f_{\rm scat} <
0.5\%$ show smaller [\ion{O}{4}]-to-12 $\mu$m ratios in average than normal
Seyferts. Next, we apply it for 23 U/LIRGs in the GOALS sample. We find
that most of the AGN-important mid- to late-stage mergers contain buried
AGNs, while the earlier stage mergers contain few. These results suggest
that the fraction of buried AGNs in U/LIRGs increase with merging
stage. Our method is applicable to individual objects whose good X-ray
spectra are not available.

\vspace{0.2in} 
We thank the reviewer for the useful comments, which helped us improve the quality of the manuscript. 
Part of this work was financially supported by the Grant-in-Aid for Scientific Research 17K05384 (Y.U.) and 15K05030 (M.I.).
This work was also supported by the Grant-in-Aid for JSPS Research Fellow 17J06407 (A.T), 17J09016 (T.K), and 18J01050 (Y.T). We acknowledge
financial support from the Ministry of Science and Technology of Taiwan
(\textit{MOST} 105-2112-M-001-029-MY3; Y.T.).  This research has made
use of data from the NASA/IPAC Infrared Science Archive and NASA/IPAC
Extragalactic Database (NED), which are operated by the Jet Propulsion
Laboratory, California Institute of Technology, under contract with the
National Aeronautics and Space Administration.

\bibliographystyle{aasjournal.bst}
\bibliography{reference}

\end{document}